\documentclass{article}
\usepackage{spconf,amsmath,graphicx}
\usepackage{listings}
\usepackage{multirow}
\usepackage{arydshln}

\usepackage{hyperref}
\let\svthefootnote\thefootnote
\newcommand\freefootnote[1]{%
  \let\thefootnote\relax%
  \footnotetext{#1}%
  \let\thefootnote\svthefootnote%
}

\title{Zero-shot Domain-sensitive Speech Recognition \\with prompt-conditioning fine-tuning}
\name{Feng-Ting Liao\textsuperscript{\text{*}}, Yung-Chieh Chan\textsuperscript{\text{*}\textdagger}, Yi-Chang Chen\textsuperscript{\textdaggerdbl}, Chan-Jan Hsu, Da-shan Shiu}
\address{MediaTek Research}

\copyrightnotice{979-8-3503-0689-7/23/\$31.00~\copyright2023 IEEE}
\begin{document}
%
\maketitle
\begin{abstract}
In this work, we propose a method to create domain-sensitive speech recognition models that utilize textual domain information by conditioning its generation on a given text prompt. This is accomplished by fine-tuning a pre-trained, end-to-end model (Whisper) to learn from demonstrations with prompt examples. We show that this ability can be generalized to different domains and even various prompt contexts, with our model gaining a Word Error Rate (WER) reduction of up to 33\% on unseen datasets from various domains, such as medical conversation, air traffic control communication, and financial meetings. Considering the limited availability of audio-transcript pair data, we further extend our method to text-only fine-tuning to achieve domain sensitivity as well as domain adaptation. We demonstrate that our text-only fine-tuned model can also attend to various prompt contexts, with the model reaching the most WER reduction of 29\% on the medical conversation dataset.

\freefootnote{\textsuperscript{\text{*}}Equal contribution. \textsuperscript{\textdagger}Work done during internship at MediaTek Research. \textsuperscript{\textdaggerdbl} Project lead. Correspondence to ft.liao@mtkresearch.com}
\freefootnote{Code available at \url{https://github.com/mtkresearch/clairaudience}}


\end{abstract}
%

\section{Introduction}
\label{sec:intro}

In the field of Automatic Speech Recognition (ASR), there has been a rapid deployment of ASR systems to downstream applications in a multitude of industry sectors, such as finance, healthcare, and transportation \cite{schulte_automatic_2020, nigmatulina_improving_2021,del_rio_earnings-22_2022}. Recent work shows the potential of improving existing ASR systems with identified domain scenarios \cite{dingliwal_domain_2022, nigmatulina_improving_2021}. Domain-Prompts \cite{dingliwal_domain_2022} tackles the challenge by learning a small set of domain-specific parameters to rescore hypotheses generated by a multi-domain ASR system whose inputs include domain names. However, the strategy is challenging to scale to unseen domains due to the resource-intensive requirements of retraining or maintaining copies of fine-tuned models.

Meanwhile, in the natural language processing (NLP) field, pre-trained large language models have exhibited excellent performance in various NLP tasks through conditional prompts \cite{brown_language_2020, openai_gpt-4_2023, wei_finetuned_2021}. A recent study underscores the efficacy of the prompting method where a text-classification model fine-tuned with prompt learning allows for notable performance on datasets in unseen domains at inference time \cite{ben-david_pada_2022}. Such an approach is appealing as it increases the number of application domains without necessitating retraining. Concretely, PADA \cite{ben-david_pada_2022} fine-tunes a T5 model \cite{raffel_exploring_2020} with prompts and domain scenarios such that at inference time, the model can utilize prompt-generated hypotheses on unseen domains for sentiment classification \cite{ben-david_pada_2022}.

Inspired by the progress in NLP, in this work, we aim to develop a zero-shot domain-sensitive speech recognition model that utilizes domain-specific information without training on the target domain. Specifically, we formulate the inclusion of the domain scenarios and tags, information easily obtainable in applications, into a prompting task. We then fine-tune the \textbf{Whisper} model, a large-scale pre-trained ASR model \cite{radford_robust_2022}, with condition prompts. Through such prompt-conditioning fine-tuning process, Whisper becomes sensitive to the given prompts and utilizes available domain information effectively. We select the Whisper model because it capitalizes rich Internet sources of paired audio-transcript data and is trained with a simple encoder-decoder architecture \cite{vaswani_attention_2017} which has been shown to scale well and flexible for including prompts \cite{radford_language_2018,raffel_exploring_2020}. During the Whisper's training phase, the Whisper decoder has processed around 6B tokens over a diverse set of web transcripts, making the decoder a robust language model for the ASR system. We thus postulate that the Whisper model has enough language ability to be further fine-tuned for becoming domain-sensitive. We call Whisper models fine-tuned with our text-prompts - \textbf{Clairaudience}. We show Clairaudience fine-tuned on Gigaspeech dataset \cite{chen_gigaspeech_2021} exhibits domain sensitivity in leveraging prior information in Medical \cite{figure_eight_inc_medical_2019}, UWB-ATCC (ATC) \cite{smidl_air_2019},  ATCO2 \cite{szoke_detecting_2021}, Earnings-22 \cite{del_rio_earnings-22_2022} datasets. Empirical results on these domain-specific datasets present a consistent reduction in terms of Word Error Rate (WER) from 17.4\% to 33.0\%.


Furthermore, acquiring large quantities of labeled data may be challenging, and only unlabelled text is available for application domains \cite{tsunoo_residual_2022, deng_adaptable_2023}. We showcase that rendering Whisper domain sensitive using text-only data with prompt-conditioning fine-tuning is achievable. We dub models fine-tuned with the approach - \textbf{Clairaudience-text-only}. Experimental outcomes of Clairaudience-text-only fine-tuned on Gigaspeech text demonstrate that the model shows no forgetting to attend to audio features and yields zero-shot improvements over the selected domains. The effectiveness of the text-only results primes us to investigate whether the model can be improved by fine-tuning on domain-text directly. This setting is most similar to the conventional domain adaptation. We find that Whisper fine-tuned on domain text achieves WER similar to but lower than Clairaudience over the ATC dataset and the splited Medical dataset.

Below we summarize our contribution:
\begin{itemize}
    \item We address the challenge of incorporating domain-specific information, such as scenarios and tags, into the Whisper model during inference without retraining on unseen domains.
    \item We propose to enable a domain-sensitive Whisper by fine-tuning the model with (i) audio-transcript pairs or with (ii) text-only data on the GigaSpeech dataset with text prompts that include domain tags generated by GPT3.5 \cite{brown_language_2020}.
    \item Our experiments show that models fine-tuned with text prompts are zero-shot domain-sensitive and showcase significant WER improvement over medical, finance, and transportation datasets. 
\end{itemize}


\section{related work}
Our work relates to several broad research areas, including prompt-engineering, multi-domain ASR, contextual-biasing, and text-only domain adaptation. In the applications to speech processing systems, prompt-engineering has only attracted attention lately. Work in \cite{peng_prompting_2023} prompts a frozen Whisper to unseen tasks, such as audio-visual recognition, code-switched speech recognition, and speech-translation. SpeechPrompt \cite{chang_speechprompt_2022} applies a fine-tuned wave2vec2 model to produce prompts as inputs to GPT2 for audio classification. Additionally, text-only training has been extensively studied in domain adaptation. Work of \cite{liu_domain-aware_2021} fine-tunes an external LM at target domains for re-scoring the outputs of ASR systems to achieve domain adaptation. Shallow fusion linearly interpolates results between an external LM and an ASR system \cite{chorowski_attention-based_2015}. In works employing contextual biasing, trained light-weight adapters embed contextual phrases to bias the embedding process in models, thereby enhancing models' ability to recognize rare words \cite{xu_cb-conformer_2023, sathyendra_contextual_2022,munkhdalai_fast_2021}. Domain-Prompts \cite{dingliwal_domain_2022}, perhaps the most similar work to ours, takes as input the audio and domain names and outputs re-scored transcription selected by a prompt-tuned GPT2 with domain-specific parameters incorporated. However, we differ from these works in that we consider the setting where multiple domains are addressed by contextually prompting a fine-tuned model with domain information.



\section{Background}
We built our approach upon the Whisper model; in this section, we provide a brief overview of the Whisper model, focusing on its connection to prompting. 

Whisper shows that ASR models trained with weakly supervised audio-transcript at scale can transfer well to existing datasets zero-shot. The Whisper model leverages encoder-decoder architecture \cite{vaswani_attention_2017}, which has shown to scale well \cite{kaplan_scaling_2020} and showcase competitive few-shot performance on tasks without fine-tuning \cite{brown_language_2020}. During training, the model is exposed to various tasks such as voice-activity detection, speech transcription, translation, and language identification. The encoder module processes log-Mel spectrograms as inputs and produces audio features. The decoder, in turn, takes these audio features along with an additional prompt token sequence to generate audio transcripts. 

The Whisper model can be conditioned to perform speech translation or transcription by specifying a series of task-specific tokens as input prompts. For instance, during inference, the token sequence can be \lstinline{<|startoftranscript|>  <|english|> <|transcribe|>} for speech transcription, while \lstinline{<|startoftranscript|>  <|german|> <|translate|>} can be employed for speech translation from German to English. During the pre-training, the model was trained with history text prepended as context to resolve ambiguous audio with long-range transcription. We thus hypothesize that, by feeding custom prompts as context text to a pre-trained Whisper, we can condition the model to leverage domain scenarios/tags.



\begin{figure*}[ht] 
\centering
\includegraphics[width=0.90\textwidth]{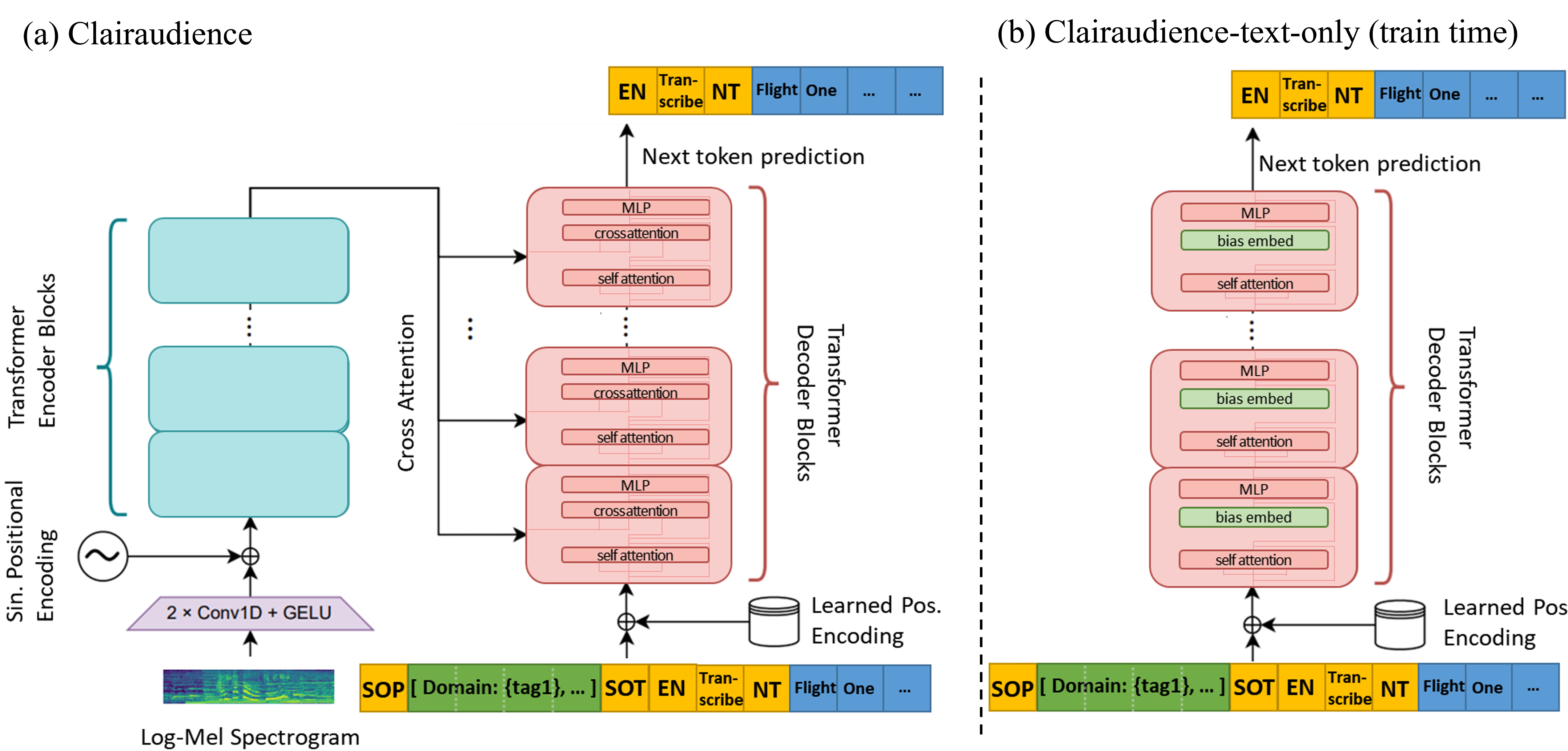}
\caption{\textbf{Cliaraudience and Clairaudience-text-only}.
\textbf{(a)} During the train time of the Clairaudience, the model takes in audio-text pairs with text having domain prompts included. The model carries out next-token prediction of the audio transcript. During the inference time, the model takes in audio-prompt pairs where each prompt includes generic domain tags from target domains. The model then generate transcripts autoregressively. \textbf{(b)} During the train time of the Clairaudience-text-only, only the transformer decoder is trained. The outputs of the cross attention at decoder layers are replaced with learnable bias terms to represent audio features. During the inference time, the bias terms are turned off, and cross-attention outputs are used instead. The inference process of Clairaudience-text-only is the same as that of Clairaudience and Whisper. Figure adapted from \cite{radford_robust_2022}.}
    \label{fig:text-only-figure}
\end{figure*}

\section{Prompt-Conditioning Fine-Tuning}
\label{prompt-conditioning fine-tuning}
Our objective is to fine-tune an end-to-end speech recognition model to have the ability to utilize the information provided by a text prompt to reduce the WER of the generated transcript. We expect the fine-tuned model will develop the capability to differentiate between phrases with similar pronunciations and recognize domain-specific words by conditioning on the text prompt. In this paper, we focus on text prompts that contain comprehensive descriptions of the broader context, scenario, and genre of the speech, information not captured in the audio clip. Our approach can be easily extended to accommodate other prompt formats, contexts, and even other modalities of prompts. 

Due to the absence of publicly available speech recognition datasets associating audio clips with text prompts, we have taken the initiative of generating this dataset ourselves. An ideal approach will be to augment Whisper's original training data in order to avoid confounding our findings with the improvement from training on extra data. As the dataset is not released, we decided to use GigaSpeech \cite{chen_gigaspeech_2021}, a dataset also collected from web audio and exhibiting rich diversity in terms of speech context. We then generate concise and diverse domain tags based on the audio transcripts, resembling extensive work in large language models (LLM) for generating data for supervision tasks \cite{taori_stanford_2023, eldan_tinystories_2023, gilardi_chatgpt_2023}. For each full audio transcript, we query GPT3.5 to generate ten domain tags in line with domain categories inherently available in the Gigaspeech dataset \cite{chen_gigaspeech_2021}.

With flexibility and extensibility in mind, we decided on inserting the prompt between the \verb!<|prev|>! token and \verb!<|startoftranscript|>! token with the format \lstinline!<|prev|>[ domain: {DOMAIN_1}, {DOMAIN_2}, ...]<|startoftranscript|>! following by the rest of decoder input sequence, where \verb!{DOMAIN_n}! represent the domain tags (see Figure \ref{fig:text-only-figure}). During training, the domain tags are sampled and incorporated into prompt text added to the decoder input.

\section{Text-only training}
\label{text-only trainings}

Whisper follows the encoder-decoder transformer architecture. At each decoder block, cross attention is performed between the layer input and the encoder output embedding, which depends on the audio clip. We remove this attention mechanism by replacing the output of each cross-attention layer with trainable biases, as illustrated in Figure \ref{fig:text-only-figure}. The addition of bias term is inspired by classifier-free guidance \cite{ho_classifier-free_2022}. The model is then tuned with conventional sequence-to-sequence training.

As the model is fine-tuned with text-only data, the model may be prone to catastrophic forgetting \cite{kirkpatrick_overcoming_2017, meng_internal_2022}. Prior work \cite{meng_internal_2022} addresses this issue through a two-stage process that initially utilizes audio-text fine-tuning, subsequently transitioning to text-only fine-tuning. However, we found that pre-trained Whisper is robust to catastrophic forgetting when the model is prompt-conditioning fine-tuned on text-only data (See Section \ref{prompt conditioning tuning experiment}). Therefore, we conduct all our text-only fine-tuning without interleaving any audio-text fine-tuning.







\begin{table*}[ht]
    \centering
    \begin{tabular}{|c|cc|c:cccc|}
    \hline
        \multirow{2}{*}{\textbf{Model}} & \multicolumn{2}{|c|}{\textbf{Domain Prompts}} & \textbf{General Domain}  &  \multicolumn{4}{c|}{\textbf{Specific Domain}}\\ 
        & train&inference &GigaSpeech-Test&Medical & ATC &ATCO2 & Earnings-22\\
        \hline
        \multirow{2}{*}{Whisper} & \multirow{2}{*}{No} &No& 
        12.36 & 8.79 & 73.82 & 42.93 & 17.86
        \\ 
        && Yes & 
        \underline{18.88} & 6.61 & 56.04 & 45.76 & \underline{26.38}
        \\ \hline
        \multirow{2}{*}{Clairaudience} & \multirow{2}{*}{Yes}&No 
        &
        11.28 & 7.29 & 61.44 & 35.73 & 17.49
        \\
        && Yes & 
        \textbf{9.22} & 6.54 & \textbf{52.47} & \textbf{28.77} & \textbf{14.75}
        \\ \hline
        
        \multirow{2}{*}{Clairaudience-text-only} & \multirow{2}{*}{Yes}&No & 
        11.70 & 7.55 & 70.45 & 42.11 & 16.69
        \\
        && Yes & 
        \underline{17.04} & \textbf{6.25} & 56.70 & 41.07 & \underline{22.93}
        \\ \hline
    \end{tabular}
    \caption{Comparison between not-tuned, audio-text tuned, and text-only tuned Whisper on prompt conditioning transcribing. Note that GigaSpeech and Earnings-22 are long-form audio datasets; Medical, ATC, ATCO2 are short-form audio datasets. The underlined scores mark the degraded performance on long-form datasets, further discussed in Section \ref{speech_naturalness}.}
    \label{table:prompt-result2}
\end{table*}

\begin{table*}[ht]
    \centering
    \begin{tabular}{|c|cc|c:c|}
    \hline
        \multirow{2}{*}{\textbf{Model}} & \multicolumn{2}{|c|}{\textbf{Domain Prompts}} & \textbf{General Domain}  &  \multicolumn{1}{c|}{\textbf{Specific Domain}}\\ 
        & train&inference &GigaSpeech-Test&Earnings-22\\
        \hline
        \multirow{1}{*}{Whisper} 
        & \multirow{1}{*}{No} 
        & Yes & 
        11.61 & 16.34
        \\ \hline
        \multirow{1}{*}{Clairaudience-text-only} & \multirow{1}{*}{Yes}
        & Yes & 
        11.67 & 15.84
        \\ \hline
    \end{tabular}
    \caption{Evaluation of not-tuned and text-only tuned Whisper after pre-pending 0.5 seconds of empty audio pads to input audio features. The padding is for simulating speech naturalness \cite{liu_how_2022} to correct errors due to segmented audio clips from long-form audio datasets.}
    \label{table:speech_naturalness}
\end{table*}

\section{EXPERIMENTAL SETUP}


\subsection{Evaluation Details}
\label{evaluation_details}

Our experiments are conducted using Whisper-large-v2 (1550M parameters)\footnote{https://github.com/openai/whisper}, the largest version of the Whisper model family,  as the base model and comparison baseline.

We generate transcripts with three beams on top candidates for reporting all the results in the paper. Following Whisper's design, we indicate the start of prediction with  \lstinline!<|startoftranscript|>! token and specify the task as English transcribing by setting the language token to \lstinline!<|english|>! and the task token to \lstinline!<|transcribe|>!. A \lstinline!<|notimestamps|>! token is then included to indicate that timestamp prediction is not required. If a text prompt is given to the decoder, it will be pre-pended to the \lstinline!<|startoftranscript|>! token with a \lstinline!<|prev|>! token in the front. Below is an example of the decoder input during inference time: \lstinline!<|prev|> [ domain: Finance, Real Estate, Investment] <|startoftranscript|>  <|english|> <|transcribe|>!.

The performance is measured by Word Error Rate \% (WER) and Word Error Rate Reduction \% (WERR) over the baseline system. 

\subsection{Data Preparation}
Two types of datasets are required in our experiments: (i) a generic dataset that resembles Whisper's training data for fine-tuning with prompts and (ii) domain-specific datasets where domain tags can be extracted or assigned for evaluating the zero-shot performance of the models. 

As described in Section \ref{prompt-conditioning fine-tuning}, we perform prompt-conditioning fine-tuning on the training set of GigaSpeech Medium (1000 hours) with GPT3.5 generated domain tags. We observe loss convergence when the model is trained on 100 hours of audio, so the amount of data required for tuning is much lower than that for pre-training an end-to-end ASR system.

To evaluate the domain sensitivity of the tuned model, we select datasets from three domains: medical conversation, air traffic control communication, and financial meetings. For medical conversation, we choose the "Medical Speech, Transcription, and Intent" (Medical) \cite{figure_eight_inc_medical_2019} dataset from Kaggle, which contains 8.5 hours of audio-text pairs and additional domain tags for each pair. The domain tags are used for constructing text prompts in the domain prompt experiment. For air traffic control, we use UWB-ATCC (ATC) \cite{smidl_air_2019} and the subset of ATCO2 \cite{szoke_detecting_2021} annotated with airport information, each containing 10 and 2 hours of clips of air traffic control communication, respectively. For financial meetings, we use Earnings-22 \cite{del_rio_earnings-22_2022}, consisting of about 100 hours of English earning calls. We will discuss our methods of extracting and assigning prompt information for individual datasets in Section \ref{prompt conditioning tuning experiment}.

\begin{table*}[ht]
    \centering
    \begin{tabular}{|c|c|c|cc|}
    \hline

        \multirow{2}{*}{\textbf{Model}} & \multicolumn{2}{|c|}{\textbf{Domain Adaptation}} & \multicolumn{2}{c|}{\textbf{Target domain}} \\ 
        
        & \multicolumn{1}{c}{Method} & Training data & ATC & Medical-split\\
        \hline
        Whisper & - & - & 73.82 & 6.07 \\ \hline
        Whisper (DA)  & \multirow{2}{*}{Fine-tuning on target domains} & audio-text & \textbf{24.04} & \textbf{4.44} \\
        \cdashline{1-1}\cdashline{3-5}
        Whisper (DA-text-only) &  & text-only & 51.84 & 5.27 \\ \hline
        Clairaudience & \multirow{2}{*}{Prompting fine-tuned model}& audio-text & 52.47 & 5.86 \\
        \cdashline{1-1}\cdashline{3-5}
        Clairaudience-text-only & & text-only & 56.7 & 5.34 \\

        \hline
    \end{tabular}
    \caption{Comparison in WER between methods that achieve domain adaptation. Whisper (DA) and Whisper (DA-text-only) are fine-tuned Whisper models on target datasets, i.e., ATC and Medical-split. Note that we split the medical dataset into train-test sets.}

    \label{table:domain-adaptation-table}
\end{table*}

\section{Experiments and results}

\subsection{Prompt Conditioning Tuning} \label{prompt conditioning tuning experiment}
This experiment aims to investigate the generalizability of the prompt conditioning ability acquired through audio-text and text-only fine-tuning on GigaSpeech with prompts that include GPT-generated domain tags. We aim to determine if this ability can be extended to diverse data genres and different types of prompt content without further adaptation.

We compare the performance in three different settings: (i) Whisper, (ii) Clairaudience, and (iii) Clairaudience-text-only. Our models are trained with prompts, and we evaluate models with or without prompts. Note that all the tunings mentioned above are conducted using GigaSpeech, and the model has yet to be exposed to the datasets used in our evaluation.

In addition to GigaSpeech, the general domain dataset which the model is tuned upon, we evaluate the three settings on four domain-specific datasets - Medical, ATC, ATCO2, and Earnings-22 - with unique prompt content. The text prompt generated for each dataset at inference time is explained as follows:

\begin{itemize}
  \item[] \textbf{GigaSpeech-Test}: We employ the identical prompt format as during training; instead of sampling, we provide the model with all ten domain phrases generated by GPT3.5. Noted that, in this section, we use the test set of GigaSpeech, which contains audio clips from a different set of web audio than that was used for training. Therefore, the model is evaluated using text prompts never encountered during training.
  
  \item[] \textbf{Medical}: Each audio clip in the dataset is accompanied by a native symptom tag. In total, 25 concise tag phrases delineate the medical symptoms, including examples like "Cough" and "Stomach ache". Leveraging these tags, we construct the text prompts associated with each audio clip.
  
  \item[] \textbf{ATC}: As the dataset does not contain any information on individual audio, we use "air traffic control" as the domain tag included in the text prompt for all the audio clips.
  
  \item[] \textbf{ATCO2}: We construct the text prompts from the airport name, communication channel type, and airline call signs associated with each audio clip. An example of the domain content in a prompt will be "Sion airport, control channel, callsigns: Alitalia Hansa KLM..."

  \item[] \textbf{Earnings-22}: Similar to ATC, the dataset contains no information on individual audio. In this case, we use the domain phrases "shareholders meeting", "earning call", "investment webinar", and "financial news".
\end{itemize}

The results of prompt conditioning transcription are presented in Table \ref{table:prompt-result2}. We observe that Whisper before fine-tuning possesses some ability to utilize the information from the text prompt to improve its transcribing accuracy. Surprisingly, it can reduce the WER from 73.82\% to 56.04\% on ATC with a simple prompt "air traffic control". However, on GigaSpeech-Test, ATCO2, and Earnings-22, adding a prompt to a not-tuned Whisper negatively impacts its prediction. In contrast, our audio-text tuned model (Clairaudience) shows improvement on all the datasets. The model demonstrates a strong ability to utilize the given text prompt to reduce the WER of its prediction. Consistent with its performance on the GigaSpeech-Test, it achieves significant WER improvement on all the datasets. As to the text-only model (Clairaudience-text-only), it shows no catastrophic forgetting in attending to audio features. The model demonstrates a weaker ability than the audio-text-tuned model but still outperforms the prompted Whisper on Medical and ATCO2 datasets. It achieves a lower WER on the Medical dataset than the audio-text-tuned model. However, similar to the prompted Whisper, the evaluation result of Clairaudience-text-only with prompt is worse than that without prompt on GagaSpeech-Test and Earnings-22. Such performance degradation is partly attributed to segmented audio in long-form audio datasets (audio length \textgreater{} 30 seconds). On the other hand, Clairaudience is immune to the problem and achieves the lowest WER. 

Interestingly, we observe that evaluation of Clairaudience and Clairaudience-text-only without prompts shows improvement over the domain-specific datasets, even though the models are fine-tuned on a general domain dataset. We postulate that such gain is contributed by better audio feature and text feature alignment learned from the general domain dataset, which is potentially presented in specific domains. Nevertheless, the outcomes of Clairaudience indicate that using domain prompts during the inference stage persistently reduces the WER on all domain-specific datasets, when contrasted between inferences with and without prompts.


\subsection{Speech Naturalness for Segmented Audio}
\label{speech_naturalness}

Given that both prompted Whisper and Clairaudience-text-only show degraded performance when evaluated on long-form audio datasets, i.e., GigaSpeech and Earnings-22, we investigated the audio inputs and found that both models often fail to transcribe segmented audios of unfinished sentences. Audios starting with unfinished sentences result in models not outputting transcripts or hallucinating. To solve the issue, we experiment with adding 0.5-second empty audio to the beginning of the input audio. The intuition is that the padded audio simulates the speech naturalness within sentences \cite{liu_how_2022}, which we assume to be presented in Whisper's pre-training data for unfinished sentences. Results in Table \ref{table:speech_naturalness} demonstrate that the heuristic effectively corrects errors. Clairaudience-text-only even outperforms the prompted Whisper on Earnings-22, consistent with results on other domain-specific datasets.

\subsection{Text-Only Domain Adaptation}

Given the results of Clairaudience-text-only, we have demonstrated the effectiveness in prompting the text-only fine-tuned model in various domains. In another research setting similar to ours - domain-adaptation, fine-tuning models on target-domain is typical \cite{deng_adaptable_2023, dingliwal_domain_2022, tsunoo_residual_2022}. We thus investigate whether the performance of the Whisper model can be improved by applying the text-only fine-tuning process under the conventional domain adaption setting.

We adapt Whisper to air traffic control communication and medical conversation with audio-text pair fine-tuning and our text-only fine-tuning method. We split the Medical dataset into adaptation and evaluation sets with an 85:15 ratio, as the default datasets do not have train-test splits. Whisper is then fine-tuned with the audio transcript or the transcript of the adaptation set and assessed on the audio-transcript pairs of the evaluation set. 


The result is presented in Table \ref{table:domain-adaptation-table}. As expected, the fine-tuned Whisper model with audio-transcript pairs on the target domain (Whisper (DA)) achieves the lowest WER. Compared with the not-tuned Whisper model, the text-only fine-tuned model on target domains (Whisper (DA-text-only)) achieves a significant 20\% and 13\% WERR improvement on ATC and Medical-split, respectively. To our surprise, the performance of Whisper (DA-text-only) is only 0.63\% and 0.07\% in absolute WER difference to Clairaudience on ATC and Clairaudience-text-only on Medical-split, respectively. The result underscores the effectiveness of applying our prompt-conditioning fine-tuning process to a pre-trained end-to-end ASR system in achieving performant ASR results through prompts instead of the conventional domain adaptation method.




\section{DISCUSSION} 
\label{discussion}
Our paper has explored a simple question: \textit{Can a fine-tuned ASR model with prompt-conditioning leverages domain information as prompts to boost performance across unseen domains without additional retraining?} Our fine-tuned models, Clairaudience and Clairaudience-text-only, improve performance against a not-tuned Whisper model and surpass the evaluation results of a prompted Whisper on most domain-specific datasets. Further experiments on text-only domain adaptation reveal that a prompted Clairaudience model can achieve similar performance in domain adaptation to the Whisper model fine-tuned directly on target domains with text-only data.




As for limitations of our study, there is a degree of subjectivity in the selected domains for evaluation, and our evaluation method work most robustly to short-audio transcription. Our first limitation is that individual domain-specific information might have appeared in the GigaSpeech training set. In Table \ref{table:prompt-result2}, we observe improvements over the inference setting without prompts for the fine-tuned model. Such improvement could be attributed to the aforementioned data overlapping between fine-tuning and testing data. Furthermore, we observe that the performance of Clairaudicne, compared with Whisper, exhibits a decrease of up to 0.95\% in absolute WER on other general domain datasets, namely Common Voice \cite{ardila_common_2020} and LibriSpeech \cite{panayotov_librispeech_2015}. The difference could be attributed to distribution deviation in fitting the model to GigaSpeech dataset, which might not be as comprehensive and diverse as the Whisper pre-training dataset. Another limitation is that Whisper was trained on audio segments (30 seconds) but not on arbitrarily long audio inputs due to memory constraints. To handle long-form transcription, the Whisper model inserts previously generated text between \verb!<|prev|>! token and \lstinline!<|startoftranscript|>! token to inform the model of the speech that precedes the current segment. Despite wrapping our text prompt with brackets, we observe that the model tuned on text-only data inherits this mechanism and mistakes the prompt as previous text when the input audio is part of an unfinished sentence as discussed in Section \ref{speech_naturalness}. This behavior is even more presented in the not-tuned model. Interestingly, despite having never seen the setting where prompt and previous text coexisted, the model trained on audio-text data is immune to such a problem. Therefore, we hypothesize that this problem can be better solved by mixing text-only tuning with a small amount of audio-text tuning or leveraging the voice detection task in Whisper as work in \cite{bain_whisperx_2023} and leaving the study for future work.



\section{CONCLUSION}

In this paper, we present a method for fine-tuning end-to-end ASR models to achieve generalization over multiple domains by prompting the fine-tuned model with domain information. We demonstrate the efficacy of the method in models fine-tuned with audio-transcript or with text-only data. Our experimental outcome shows that: \textbf{(i)} Clairaudience, prompt-conditioning fine-tuned on GigaSpeech, is zero-shot domain-sensitive on Medical, ATC, ATCO2, and Earnings-22 datasets. It achieves the best WER in comparison to Whisper and Clairaudience-text-only; \textbf{(ii)} Whisper can be rendered domain-sensitive with text-only fine-tuning of its decoder. The resulting Clairaudience-text-only model outperforms Whisper on Medical, ATCO2, and Earnings-22 datasets; \textbf{(iii)} Through domain prompts, Clairaudience can perform similarly to Whisper (DA-text-only), which is text-only fine-tuned directly on ATC and Medical-split datasets. We hope that our paper will motivate further investigation into more prompt-based ASR systems, zero-shot learning, and using text-only data to improve end-to-end ASR systems.


\bibliographystyle{IEEEbib} %
\bibliography{paper_bib}


\end{document}